\documentclass[prb,showpacs,floatfix,twocolumn,letterpaper]{revtex4}
\usepackage{graphicx} 
\usepackage[usenames,dvipsnames]{color}

\newcommand{\nco}{NiCr$_2$O$_4$}

\begin{document}

\title {Magnetocapacitance as a sensitive probe of magnetostructural changes in \nco\,}

\author{Taylor\,D.\,Sparks}\email{sparks@eng.utah.edu}
\altaffiliation[Current Address: ] {Materials Science and Engineering Department,
University of Utah, Salt Lake City, UT, 84112}
\affiliation{Materials Department and Materials Research Laboratory\\
University of California, Santa Barbara, CA, 93106, USA}
\author{Moureen\,C.\,Kemei}\email{kemei@mrl.ucsb.edu}
\affiliation{Materials Department and Materials Research Laboratory\\
University of California, Santa Barbara, CA, 93106, USA}
\author{Phillip\,T.\,Barton}\email{pbarton@mrl.ucsb.edu}
\affiliation{Materials Department and Materials Research Laboratory\\
University of California, Santa Barbara, CA, 93106, USA}
\author{Ram\,Seshadri}\email{seshadri@mrl.ucsb.edu}
\affiliation{Materials Department and Materials Research Laboratory\\
University of California, Santa Barbara, CA, 93106, USA}
\author{Eun-Deok\,Mun}\email{edmun@lanl.gov}
\affiliation{National High Magnetic Field Laboratory\\
Los Alamos National Laboratory, NM, 87545, USA}
\author{Vivien\,Zapf}\email{vzapf@lanl.gov}
\affiliation{National High Magnetic Field Laboratory\\
Los Alamos National Laboratory, NM, 87545, USA}

\date{\today} 

\begin{abstract}
The spinel \nco\, is characterized using dielectric and high magnetic field measurements. The trends in the magnetodielectric response fall into three clear temperature regimes corresponding to known magnetic and structural transitions. Above 65\,K, weak magnetic field dependence of the dielectric constant is observed with no hysteresis. When 30\,K\,$\leq\,T\,\leq$\,65\,K, a strong dependence of the dielectric constant on the magnetic field is observed and hysteresis develops resulting in so called butterfly loops. Below 30\,K, magnetodielectric hysteresis is enhanced. Magnetodielectric hysteresis mirrors magnetic hysteresis suggesting that spin-spin interactions are the mechanism for the magnetoelectric effect in \nco. At high fields however, the magnetization continues to increase while the dielectric constant saturates. Magnetodielectric measurements of \nco\, suggest an additional, previously unobserved transition at 20\,K. Subtle changes in magnetism and structure suggest that this 20\,K anomaly corresponds to the completion of ferrimagnetic ordering and the spin driven structural distortion. We demonstrate that magnetocapacitance is a sensitive probe of magnetostructural distortion.

\end{abstract}

\maketitle

Coupling of spin and charge in insulating materials can give rise to magnetoelectric effects which enable the control of magnetic polarization using an electric field, or conversely, the reversal of electric polarization by a magnetic field. Numerous applications of the magnetoelectric effect have been identified by Wood and Austin such as spin-charge transducers, tunable filters, actuators, magnetic sensors, and multiple-state memory elements.\cite{Wood1975} However, few technologies based on the magnetoelectric effect in insulators have been realized due to the small magnitude of induced magnetic or electric polarization, weak coupling between charge and spin, and the low-temperatures where magnetoelectric coupling often arises. In fact, there is a dearth of single-phase materials exhibiting strong magnetoelectric coupling. In the last decade, intense effort has been devoted to understanding how magnetism that breaks spatial-inversion symmetry can induce ferroelectricity in so-called type-two multiferroics, especially in the REMnO$_3$ systems.\cite{kimura_2012, Kimura2003N} More recently, room temperature magnetoelectricity due to complex spin ordering has been observed in hexaferrites.\cite{Kitagawa_2010, kimura_2012} Electric field control of four different magnetoelectric states in the hexaferrite Ba$_{0.52}$Sr$_{2.48}$Co$_2$Fe$_{24}$O$_{41}$ reveals new opportunities for the next-generation of memory devices based on magnetoelectric materials.\cite{chun_2012}

Advancing the understanding of known magnetodielectric compounds is an important challenge that will guide the design of future multiferroic materials. The normal spinel \nco\, is a known magnetodielectric that was first reported by Mufti $et\,al.$\cite{crottaz_1997,mufti_2010} \nco\, crystallizes in the cubic space group $Fd\overline{3}m$ above 310\,K.\cite{crottaz_1997} Ni$^{2+}$ ions have the degenerate electronic configuration $e^4\,t_2^4$ in tetrahedral coordination while non-degenerate Cr$^{3+}$ $e_g^3$ prefer to occupy octahedral sites. Below 310\,K, orbital degeneracy on tetrahedral Ni$^{2+}$ drives cooperative Jahn-Teller distortion of \nco\, resulting in the lowering of average structural symmetry from cubic $Fd\overline{3}m$ to tetragonal $I4_1/amd$.\cite{dunitz_1957,kocsis_2013} This tetrahedral distortion results in the elongation of NiO$_4$ tetrahedra and $c/a$\,$>$\,1. Magneto-structural coupling drives further distortion of \nco\, from tetragonal symmetry to orthorhombic symmetry at the N\'eel temperature ($T_N$\,=65\,K).\cite{Suchomel2012PRB,ishibashi_2007} A second distortion within the orthorhombic structure takes place at $T$\,=\,30\,K where anomalies in magnetic susceptibility and heat capacity are observed.\cite{Suchomel2012PRB,tomiyasu_2004,Klemme_2002,kocsis_2013} 

Neutron diffraction studies by Kagomiya and Tomiyasu show that the ordering of the longitudinal ferrimagnetic component of \nco\, occurs at 60\,K and this is followed by the ordering of the transverse antiferromagnetic components at 30\,K.\cite{tomiyasu_2004} Neutron scattering reveals four Cr$^{3+}$ $B$ sublattices, that give rise to a net moment of 2.69\,$\mu_B$ along the [100] direction, this moment is compensated by the net $A$ sublattice moment of 3.0\,$\mu_B$ resulting in an overall moment of 0.31\,$\mu_B$ $per$ formula unit of \nco.\cite{tomiyasu_2004} Detailed heat capacity measurements by Klemme and Miltenburg show three anomalies: at 310\,K due to Jahn-Teller ordering, near 70\,K and at 29\,K due to magneto-structural coupling.\cite{Klemme_2002} Mufti $et\,al.$ have reported changes in the slope of the dielectric constant of \nco\, at 75\,K and 31\,K; they also show magnetic field dependence of the dielectric constant of \nco.\cite{mufti_2010} Following this initial work, Maignan $et\,al.$ measured a polarization of 13\,$\mu$C\,m$^{-2}$ in the ferrimagnetic state of \nco.\cite{maignan_2012} Polarization in the ferrimagnetic state of the related spinels FeCr$_2$O$_4$ and CoCr$_2$O$_4$ has also been observed.\cite{singh_2011}

Here, we present a carefully study of the temperature and magnetic field-dependence of polycrystalline \nco\,, revealing three temperature regimes that describe the trends in magnetocapacitance. We provide the first high-field magnetization measurements of the spinel \nco\, at various temperatures. We describe the correlations between spin-spin interactions and the dielectric constant of \nco. We also discuss the multiferroic properties of \nco\, in the context of recent studies of magnetostructural coupling in \nco\, showing that variations in magnetocapacitance occur concurrently with changes in structure and spin configuration.\cite{Suchomel2012PRB}  We illustrate the sensitivity of magnetocapacitance measurements to magnetostructural changes in \nco\, by revealing an unreported anomaly at 20\,K. Incorporating detailed magnetic and structural studies, we investigate the origin of the anomaly at 20\,K. We conclude that the 20\,K anomaly corresponds to the completion of structural and magnetic ordering. 

\begin{figure}
\centering
\includegraphics[scale=0.9]{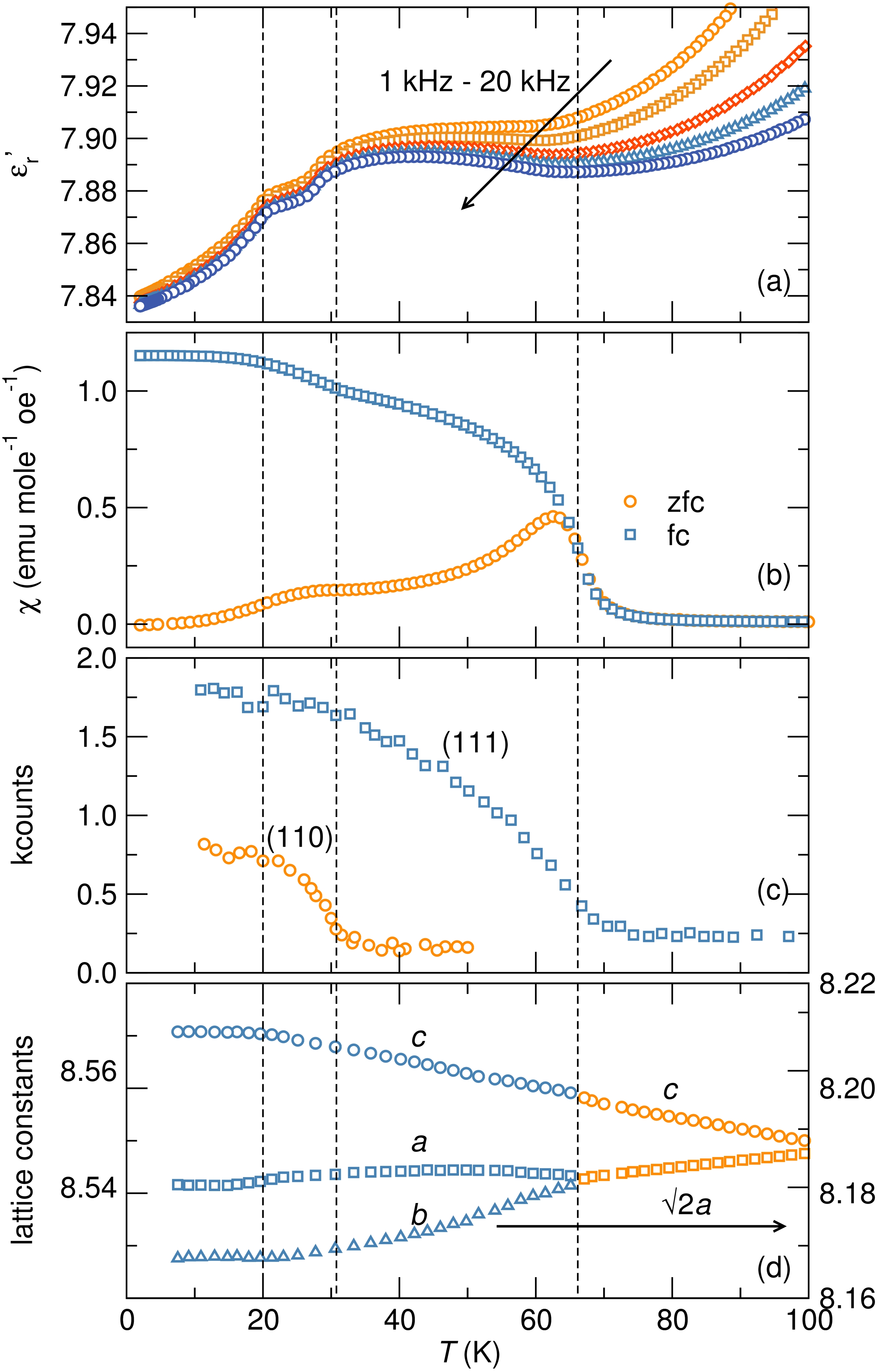}
\caption{Magnetostructural and magnetodielectric coupling in \nco. (a) Temperature-dependent dielectric measurements of \nco\, shows anomalies at 65\,K and 30\,K due to magnetic ordering. A previously unreported anomaly is observed at 20\,K. (b) Zero field cooled (ZFC) and field cooled (FC) temperature-dependent magnetic susceptibility of \nco\, reveal the onset of ferrimagnetic order at 65\,K where the ZFC and FC curves diverge. At 30\,K, another change in the magnetic structure occurs. We note that at 20\,K, the ZFC and FC curves saturate. Neutron diffraction studies of \nco\, reported by Tomiyasu and Kagomiya are shown in (c).\cite{tomiyasu_2004} The fundamental (111) reflection increases in intensity below 65\,K while the superlattice (110) reflection increases in intensity below 30\,K. The neutron reflections attain stable values below 20\,K. (d) Lattice parameters of \nco\, as a function of temperature; the tetragonal to orthorhombic transition occurs at 65\,K followed by further structural change at 30\,K within the orthorhombic $Fddd$ spacegroup. Below 20\,K, the lattice constants plateau.}
\label{fig:temp}
\end{figure}

\section{Methods}

Phase pure powders of NiCr$_2$O$_4$ were prepared from stoichiometric amounts of NiO and Cr$_2$O$_3$ that were ground, pelletized, and annealed at 800\,$^{\circ}$ for 12 hours. Sample pellets were reground, repelletized, and annealed at 1100\,$^{\circ}$ for 24 hours.  Structural characterization was performed using high-resolution ($\Delta Q/Q\,\leq\,2 \times 10^{-4}$) synchrotron X-ray powder diffraction at the Advanced Photon Source, Argonne National Laboratory. Powder patterns were fit to the crystal structure using the Rietveld refinement method as implemented in the EXPGUI/GSAS program.\cite{toby_expgui_2001,larson_2000} Magnetic susceptibility measurements were performed using a Quantum Design (QD) Magnetic Property Measurement System. High field magnetization was measured using an extraction magnetometer in a capacitor-bank-driven 65 Tesla pulsed-field magnet at the National High Magnetic Field Laboratory Pulsed-Field Facility at Los Alamos National Laboratory.\cite{detwiler_2000} Samples for magnetocapacitance measurements were spark plasma sintered at 1200\,K under a load of 6\,kN for 10 minutes with a very fast heating and cooling cycle. Spark plasma sintering was used to achieve high densities ensuring the collection of highly reliable dielectric measurements. Densified pellets were annealed in air to 1100\,K for 3 hours to ensure stoichiometric \nco\, is recovered following slight reduction during the spark plasma sintering process. Laboratory Cu$K\alpha$ X-ray diffraction performed on annealed spark plasma sintered pellets show that stoichiometric \nco\, is obtained.

Dielectric measurements as a function of temperature and applied magnetic field were carried out using an Andeen-Hagerling AH2700A capacitance bridge connected via shielded coaxial cables to a sample located within a QD Dynacool Physical Property Measurement System. Coplanar faces of a cylindrical sample 9.5\,mm in diameter were polished and coated with conducting epoxy to enable the connection of electrical contacts. The sample was clamped in place to restrict movement within the magnetic field. The sample was 2.3\,mm thick. Capacitance measurements were collected at several frequencies (1\,kHz, 2\,kHz, 5\,kHz, 10\,kHz and 20\,kHz) continuously as the temperature or magnetic field was varied at 3\,K/min or 150\,Oe/s, respectively. 

\section{Results}

Synchrotron X-ray powder diffraction of \nco\, collected at 100\,K shows that the prepared sample is well modelled by the tetragonal spacegroup $I4_1/amd$ with lattice parameters $a\,=\,$5.79183(2)\,\AA\, and $c\,=\,$8.53835(5)\,\AA. The lattice parameters reported here are in good agreement with values reported in the literature.\cite{Suchomel2012PRB} Oxygen atoms are described by the general positions $x\,=0\,$,  $y\,=\,0.5129(2)$ and $z\,=\,0.2329(1)$ while Ni$^{2+}$ and Cr$^{3+}$ occupy special positions (0,$\frac{1}{4}$,$\frac{3}{8}$) and (0,0,0) respectively. \\

Temperature-dependent dielectric measurements of \nco\, show anomalies at 20\,K, 30\,K, and 65\,K [Fig.\,\ref{fig:temp} (a)]. These anomalies correspond to structural and magnetic transitions in NiCr$_2$O$_4$.  Ferrimagnetic ordering occurs at 65\,K as shown by divergence of zero-field cooled and field cooled temperature-dependent susceptibility measurements shown in Fig. \ref{fig:temp} (b).  In agreement with the susceptibility studies, neutron diffraction studies by Tomiyasu and Kagomiya show the enhancement of the fundamental (111) reflections below 65\,K due to the the ordering of the longitudinal ferrimagnetic component of \nco\, [Fig. \ref{fig:temp} (c)].\cite{tomiyasu_2004} Further change in the magnetic structure occurs at 30\,K where another anomaly in the temperature-dependent susceptibility is observed [Fig. \ref{fig:temp} (b)] and (110) superlattice reflections emerges in neutron scattering studies as a result of ordering of the transverse antiferromagnetic component of \nco\, [Fig. \ref{fig:temp} (c)].\cite{tomiyasu_2004} Magnetic ordering in \nco\, is strongly coupled to structure as reported by Suchomel $et\,al.$\cite{Suchomel2012PRB} At 65\,K, ferrimagnetic ordering is accompanied by a tetragonal $I4_1/amd$ to orthorhombic $Fddd$ structural distortion [Fig. \ref{fig:temp} (d)]. Further change in magnetic ordering at 30\,K also occurs concurrently with further structural distortion of \nco\, within the orthorhombic $Fddd$ spacegroup [Fig. \ref{fig:temp} (d)]. \\

At 20\,K, a clear anomaly is observed in magnetodielectric measurements of \nco [Fig. \ref{fig:temp} (a)]. Reevaluation of temperature-dependent magnetic susceptibility, neutron reflection intensities, and lattice constants reveals signatures of structural and magnetic changes at 20\,K (Fig. \ref{fig:temp}). Notably, all the temperature-dependent structural and magnetic parameters attain steady values at 20\,K.  These trends in neutron intensity, magnetic susceptibility and lattice parameters suggest that magnetic and structural ordering in \nco\, occurs over a wide temperature range, between 65\,K and 20\,K, finally reaching completion at 20\,K. This finding of the continuous change in spin and lattice structure between 65\,K and 20\,K  is facilitated by high precision magnetocapacitance measurements. \\

\begin{figure}
\centering
\includegraphics[scale=0.75]{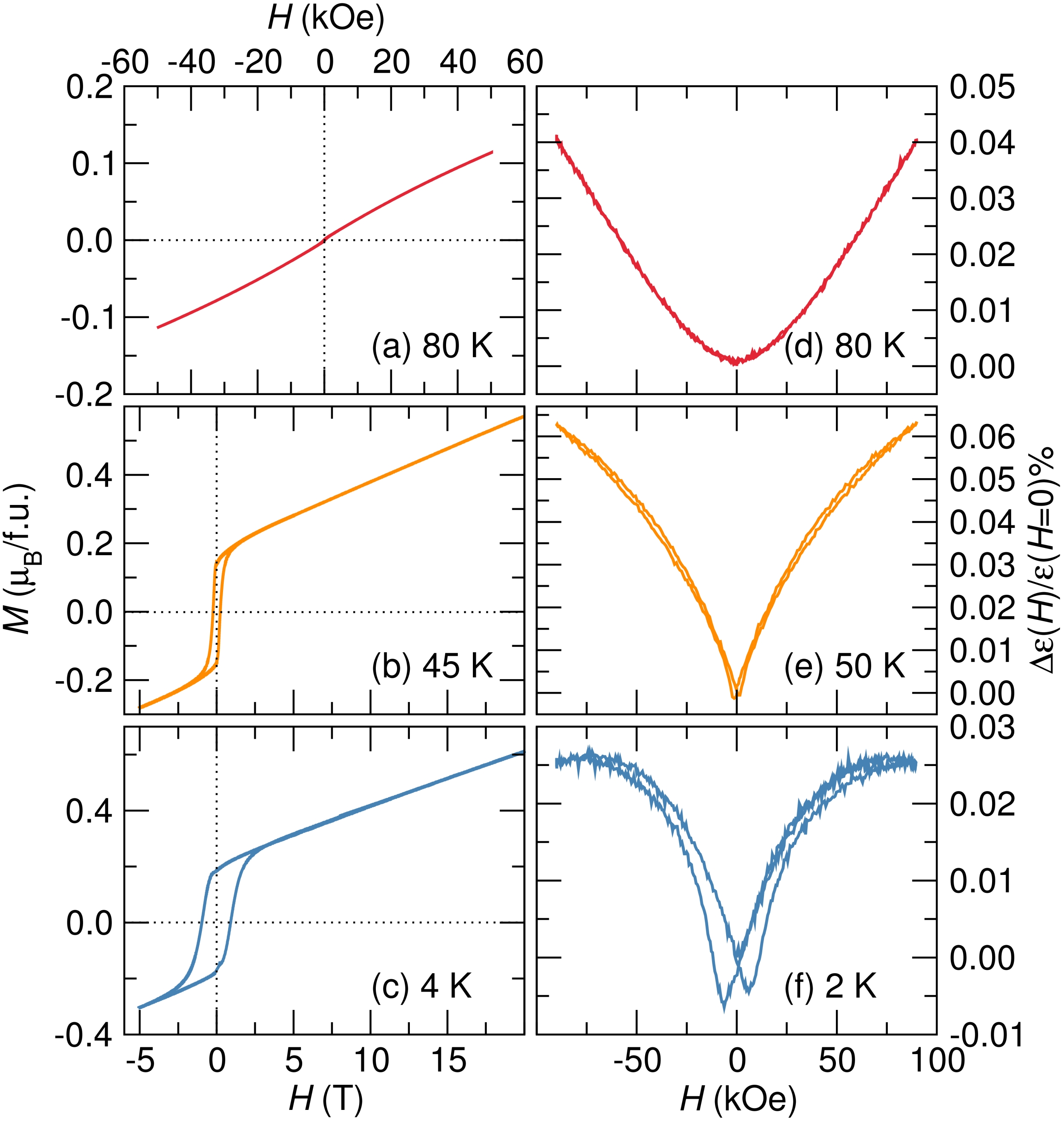}
\caption{Magnetization and magnetodielectric trends in \nco. Field-dependent magnetization measurements collected at (a) 80\,K, (b) 45\,K and (c) 4\,K are shown alongside percentage changes in the dielectric constant in a varying magnetic field measured at (d) 80\,K, (e) 50\,K, and (f) 2\,K. Above the N\'eel temperature, there is no hysteresis in the magnetization or in the field-dependent dielectric constant [(a) and (d)]. Below the N\'eel temperature, slight hysteresis is observed in both field-dependent magnetization and dielectric measurements [(b) and (e)]. Below the second magnetic transition at 30\,K, hysteresis is further enhanced in both magnetization and dielectric measurements. High field measurements collected at 4\,K (b) and 45\,K (c) show the suppression of hysteresis above 2.5\,T. A linear increase in magnetization is observed with increase in field upto 60\,T (Data above 20\,T is not shown for better illustration of the hysteresis at low fields).}
\label{fig:H}
\end{figure}

A linear dependence of the magnetization on the applied field is observed above the N\'eel temperature as shown in Fig. \ref{fig:H} (a). Coercivity develops below the ferrimagnetic ordering temperature of \nco\,[Fig. \ref{fig:H} (b)].  Further enhancement of the coercivity is observed below the second magnetic ordering temperature of 30\,K as illustrated in Fig. \ref{fig:H} (c). The highest coercivity of $\sim$\,8.3\,kOe is observed near 4\,K. High field magnetization studies performed at 4\,K and at 45\,K reveal a suppression of hysteresis and a linear increase of magnetization above 2.5\,T [Fig. \ref{fig:H} (b) and (c)]. At 4\,K and 60\,T, the magnetization of \nco\, is about 1.6\,$\mu$B $per$ formula unit. This magnetization value is 2.4\,$\mu_B$ less than the expected value for a collinear ferrimagnetic configuration in \nco\, suggesting that the magnetic structure of \nco\, reported by Kagomiya and Tomiyasu, which consists of longitudinal and transverese magnetic sublattices, persists to high fields with slight canting of spins contributing to the linear enhancement of the magnetization with field.\cite{tomiyasu_2004} \\

Isothermal field-dependent dielectric measurements exhibit trends that are well described by three temperature regimes [Fig. \ref{fig:H} (d), (e), and (f)]. Above 65\,K [Fig. \ref{fig:H} (d)], no hysteresis is observed and the field dependence is parabolic with $\Delta \varepsilon(H)/\varepsilon(H=0)$\% reaching 0.036\,\%. In the temperature range 30\,K\,$\leq\,T\,\leq$\,65\,K [Fig. \ref{fig:H} (e)],  hysteresis develops and $\Delta \varepsilon(H)/C(H=0)$\% reaches the maximum value observed in \nco\, of 0.065\,\%. The strong-dependence of the dielectric constant on the applied field between 30\,K and 65\,K becomes weaker when $T\,\leq\,$30\,K while strong hysteresis emerges in the field-dependent dielectric response [Fig. \ref{fig:H} (f)]. As we discuss below, the trends in the three regimes follow closely the field dependence of the squared magnetization.

To ensure that the observed magnetoelectric response in \nco\, is not merely due to magnetoresistance (Maxwell-Wagner effects) we examine the frequency and field dependence of the dielectric constant and loss. \cite{catalan_2006} Frequency dependence of the dielectric constant is observed above 80\,K illustrating that resistive artifacts affect the dielectric behavior at these high temperatures [Fig. \ref{fig:temp} (a)]. As the measurement approaches 65\,K the frequency dependence is quenched indicating  true intrinsic magnetodielectric coupling in \nco. The strong dependence of the dielectric constant on the applied field below the magnetic ordering temperature (65\,K) also illustrates the intrinsic nature of magnetocapacitance in \nco\, [Fig. \ref{fig:H} (e) and (f)]. Catalan has shown that magnetoresistive effects have signatures in the dielectric loss. The dielectric loss of \nco\, measured below the N\'eel temperature, where strong magnetocapacitance arises, is about 0.0034\,nS. The loss remains constant in a varying magnetic field while the dielectric constant strongly depends on field as shown in Fig. \ref{fig:H} (e) and (f).\cite{catalan_2006}

\section{Discussion}

Dielectric anomalies at 65\,K, 30\,K, and 20\,K occur concurrently with changes in magnetic structure [Fig. \ref{fig:temp} (a), (b), and (c)].\cite{tomiyasu_2004} This correlation supposes that magnetic perturbations couple to the dielectric constant of \nco. The coupling between spin-spin correlations and phonons in \nco\, is not surprising in light of the recent work of Suchomel $et\,al.$ which shows lattice distortions occurring concomitantly with magnetic ordering [Fig. \ref{fig:temp} (d)].\cite{Suchomel2012PRB} Similar spin-phonon mediated magnetoelectric coupling has been reported for the magnetoelectrics Mn$_3$O$_4$,\cite{tackett_2007} SeCuO$_3$,\cite{lawes_2003} and TeCuO$_3$.\cite{lawes_2003} 

\begin{figure}
\centering
\includegraphics[scale=0.8]{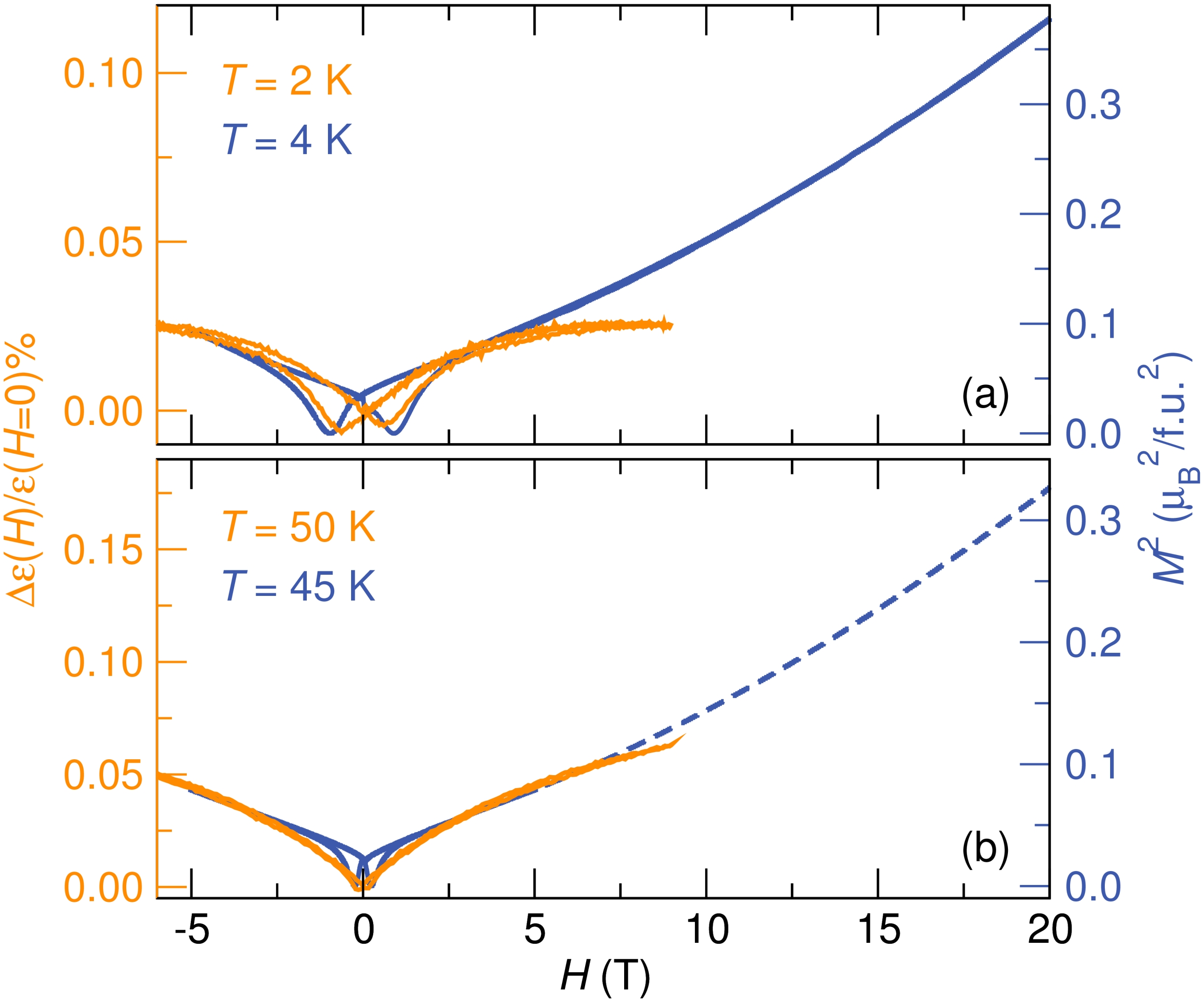}
\caption{Correlations between the field-dependent squared magnetization and the dielectric constant near 5\,K (a) and 50\,K (b). Field-dependent changes in the dielectric constant correlate with trends in the field-dependent squared magnetization at low fields. Deviations between the field-dependent dielectric constant and the squared magnetization occur at 6\,T when $T$\,$\sim$\,5\,K, and at 7.5\,T for $T$\,$\sim$\,50\,K. At high fields, the dielectric constant becomes nearly field independent approaching a steady value while the squared magnetization continuous to increase with field.}
\label{fig:CM}
\end{figure}

Changes in the squared magnetization in the presence of an applied field below 65\,K accurately depict the trends in magnetodielectric coupling at low fields as shown in Fig. \ref{fig:CM}. The relative changes in dielectric constant trace the squared magnetization near 5\,K, and below 6\,T [Fig. \ref{fig:CM} (a)]. With increase in temperature, the agreement between the squared magnetization and the relative changes in capacitance persist to higher fields; at around 50\,K, the deviation between the squared magnetization and the changes in dielectric constant begins at 7.5\,T [Fig. \ref{fig:CM} (b)]. A continuous nearly linear increase in squared magnetization occurs at high fields while the percentage changes in dielectric constant seem to level off. The agreement between the squared magnetization and the changes in dielectric constant at low fields illustrate that the magnetoelectric coupling term $\gamma P^2M^2$ of the Ginzburg-Landau theory for second-order phase transitions (Equation \ref{thermodynamicpotential}) is the most significant in \nco\, as was earlier proposed by Mufti $et\,al$.\cite{kimura_2003, mufti_2010} In the thermodynamic potential $\Phi$ expression in Equation \ref{thermodynamicpotential}, $\alpha$, $\beta$, $\alpha '$, $\beta '$ and $\gamma$ are temperature-dependent magnetoelectric coupling coefficients, $M$ and $P$ are the magnetization and polarization respectively, and $E$ and $H$ are the electric and magnetic fields respectively. We find that this spin mediated magnetodielectric coupling is most significant at low fields. 

\begin{equation}
\Phi = \Phi_0 + \alpha P^2 + \frac{\beta}{2}P^4 - PE + \alpha 'M^2 + \frac{\beta '}{2}M^4-MH + \gamma P^2M^2
\label{thermodynamicpotential}
\end{equation}

The magnetoelectric response, particularly the field dependence of capacitance, yields unique insight to subtle structural distortions in \nco. The details of the structural change at 30\,K in \nco\, are not known beyond a slight change in temperature-dependent lattice constants resulting from the elongation of NiO$_4$ tetrahedra. Nevertheless, the increased magnitude in magnetoelectric hysteresis could possibly suggest a cation off-centering that was not observed in previous structural studies. Polarization measurements by Maignan $et\,al.$ have demonstrated that \nco\, is indeed a multiferroic with polarization that develops near the N\'eel temperature and continues to increase until 20\,K where it approaches a steady value.\cite{maignan_2012}

\section{Conclusion}

In summary, the magnetodielectric response of dense, single phase \nco\, samples was measured as a function of frequency, field, and temperature. The largest magnetoelectric hysteresis was observed in the $Fddd$ phase below 30\,K while the largest $\Delta \varepsilon(H)/\varepsilon(H=0)$\%\,=\,0.06\%, was also observed in the $Fddd$ phase but in the temperature range 30\,K\,$\leq$\,T\,$\leq$\,65\,K. We discuss magnetocapacitance in the context of recently reported magnetostructural coupling in \nco. We demonstrate the coupling of the spin-spin correlation function to the dielectric constant of \nco\, which results in three temperature regimes with varying trends in magnetocapacitance. We show the agreement between the squared magnetization and the relative changes in dielectric constant below 65\,K and at low fields. We report a 20\,K anomaly in \nco\, that is evident in magnetodielectric measurements and has subtle features in structural and magnetic order parameters. This anomaly corresponds to the completion of spin and structural transformations. We show that magnetocapacitance is a sensitive probe of magnetostructural coupling in \nco. We present high-field magnetization studies of the spinel \nco\, which show a linearly increasing magnetization upto 65\,K at 45\,K and 4\,K. 

\section{Acknowledgements}
This project was supported by the NSF through the DMR 1105301. MCK is supported by the Schlumberger Foundation Faculty for the Future fellowship. PTB is supported by a National Science Foundation Graduate Research Fellowship. We acknowledge the use of shared experimental facilities of the Materials Research Laboratory: an NSF MRSEC, supported by NSF DMR 1121053. The 11-BM beamline at the Advanced Photon Source is supported by the DOE, Office of Science, Office of Basic Energy Sciences, under Contract No. DE-AC0206CH11357. The National High Magnetic Field Laboratory is supported by the National Science Foundation through Cooperative Grant No. DMR 1157490, the State of Florida, and the U.S. Department of Energy. We thank B. C. Melot for helpful discussions and assistance in designing and building the magnetocapacitance measurement system. MCK thanks David Poerschke for assistance with the spark plasma sintering process. 

\bibliography{NiCr2O4_PRB_Short}

\begin{thebibliography}{22}
\expandafter\ifx\csname natexlab\endcsname\relax\def\natexlab#1{#1}\fi
\expandafter\ifx\csname bibnamefont\endcsname\relax
  \def\bibnamefont#1{#1}\fi
\expandafter\ifx\csname bibfnamefont\endcsname\relax
  \def\bibfnamefont#1{#1}\fi
\expandafter\ifx\csname citenamefont\endcsname\relax
  \def\citenamefont#1{#1}\fi
\expandafter\ifx\csname url\endcsname\relax
  \def\url#1{\texttt{#1}}\fi
\expandafter\ifx\csname urlprefix\endcsname\relax\def\urlprefix{URL }\fi
\providecommand{\bibinfo}[2]{#2}
\providecommand{\eprint}[2][]{\url{#2}}

\bibitem[{\citenamefont{Wood and Austin}(1975)}]{Wood1975}
\bibinfo{author}{\bibfnamefont{V.~E.} \bibnamefont{Wood}} \bibnamefont{and}
  \bibinfo{author}{\bibfnamefont{A.~E.} \bibnamefont{Austin}},
  \emph{\bibinfo{title}{Magnetoelectric Interaction Phenomena in Crystals}}
  (\bibinfo{publisher}{Gordon and Breach}, \bibinfo{address}{London},
  \bibinfo{year}{1975}).

\bibitem[{\citenamefont{Kimura}(2012)}]{kimura_2012}
\bibinfo{author}{\bibfnamefont{T.}~\bibnamefont{Kimura}},
  \bibinfo{journal}{Annu. Rev. Condens. Matter Phys.}
  \textbf{\bibinfo{volume}{3}}, \bibinfo{pages}{93} (\bibinfo{year}{2012}).

\bibitem[{\citenamefont{Kimura et~al.}(2003{\natexlab{a}})\citenamefont{Kimura,
  Goto, Shintani, Ishizaka, Arima, and Tokura}}]{Kimura2003N}
\bibinfo{author}{\bibfnamefont{T.}~\bibnamefont{Kimura}},
  \bibinfo{author}{\bibfnamefont{T.}~\bibnamefont{Goto}},
  \bibinfo{author}{\bibfnamefont{H.}~\bibnamefont{Shintani}},
  \bibinfo{author}{\bibfnamefont{K.}~\bibnamefont{Ishizaka}},
  \bibinfo{author}{\bibfnamefont{T.}~\bibnamefont{Arima}}, \bibnamefont{and}
  \bibinfo{author}{\bibfnamefont{Y.}~\bibnamefont{Tokura}},
  \bibinfo{journal}{Nature} \textbf{\bibinfo{volume}{426}}, \bibinfo{pages}{55}
  (\bibinfo{year}{2003}{\natexlab{a}}).

\bibitem[{\citenamefont{Kitagawa et~al.}(2010)\citenamefont{Kitagawa,
  Y.~Hiraoka, Ishikura, Nakamura, and Kimura}}]{Kitagawa_2010}
\bibinfo{author}{\bibfnamefont{Y.}~\bibnamefont{Kitagawa}},
  \bibinfo{author}{\bibfnamefont{T.~H.} \bibnamefont{Y.~Hiraoka}},
  \bibinfo{author}{\bibfnamefont{T.}~\bibnamefont{Ishikura}},
  \bibinfo{author}{\bibfnamefont{H.}~\bibnamefont{Nakamura}}, \bibnamefont{and}
  \bibinfo{author}{\bibfnamefont{T.}~\bibnamefont{Kimura}},
  \bibinfo{journal}{Nat. Mater.} \textbf{\bibinfo{volume}{9}},
  \bibinfo{pages}{797} (\bibinfo{year}{2010}).

\bibitem[{\citenamefont{Chun et~al.}(2012)\citenamefont{Chun, Chai, Jeon, Kim,
  Oh, Kim, Kim, Jeon, Haam, Park et~al.}}]{chun_2012}
\bibinfo{author}{\bibfnamefont{S.~W.} \bibnamefont{Chun}},
  \bibinfo{author}{\bibfnamefont{Y.~S.} \bibnamefont{Chai}},
  \bibinfo{author}{\bibfnamefont{B.-G.} \bibnamefont{Jeon}},
  \bibinfo{author}{\bibfnamefont{H.~J.} \bibnamefont{Kim}},
  \bibinfo{author}{\bibfnamefont{Y.~S.} \bibnamefont{Oh}},
  \bibinfo{author}{\bibfnamefont{I.}~\bibnamefont{Kim}},
  \bibinfo{author}{\bibfnamefont{H.}~\bibnamefont{Kim}},
  \bibinfo{author}{\bibfnamefont{B.~J.} \bibnamefont{Jeon}},
  \bibinfo{author}{\bibfnamefont{S.~Y.} \bibnamefont{Haam}},
  \bibinfo{author}{\bibfnamefont{J.-Y.} \bibnamefont{Park}},
  \bibnamefont{et~al.}, \bibinfo{journal}{Phys. Rev. Lett.}
  \textbf{\bibinfo{volume}{108}}, \bibinfo{pages}{177201}
  (\bibinfo{year}{2012}).

\bibitem[{\citenamefont{Crottaz et~al.}(1997)\citenamefont{Crottaz, Kubel, and
  Schmid}}]{crottaz_1997}
\bibinfo{author}{\bibfnamefont{O.}~\bibnamefont{Crottaz}},
  \bibinfo{author}{\bibfnamefont{F.}~\bibnamefont{Kubel}}, \bibnamefont{and}
  \bibinfo{author}{\bibfnamefont{H.}~\bibnamefont{Schmid}},
  \bibinfo{journal}{J. Mater. Chem.} \textbf{\bibinfo{volume}{7}},
  \bibinfo{pages}{143} (\bibinfo{year}{1997}).

\bibitem[{\citenamefont{Mufti et~al.}(2010)\citenamefont{Mufti, Nugroho, Blake,
  and Palstra}}]{mufti_2010}
\bibinfo{author}{\bibfnamefont{N.}~\bibnamefont{Mufti}},
  \bibinfo{author}{\bibfnamefont{A.~A.} \bibnamefont{Nugroho}},
  \bibinfo{author}{\bibfnamefont{G.~R.} \bibnamefont{Blake}}, \bibnamefont{and}
  \bibinfo{author}{\bibfnamefont{T.~T.~M.} \bibnamefont{Palstra}},
  \bibinfo{journal}{J. Phys.: Condens. Matter} \textbf{\bibinfo{volume}{22}},
  \bibinfo{pages}{075902} (\bibinfo{year}{2010}).

\bibitem[{\citenamefont{Dunitz and Orgel}(1957)}]{dunitz_1957}
\bibinfo{author}{\bibfnamefont{J.~D.} \bibnamefont{Dunitz}} \bibnamefont{and}
  \bibinfo{author}{\bibfnamefont{L.~E.} \bibnamefont{Orgel}},
  \bibinfo{journal}{J. Phys. Chem. Solids} \textbf{\bibinfo{volume}{3}},
  \bibinfo{pages}{20} (\bibinfo{year}{1957}).

\bibitem[{\citenamefont{Kocsis et~al.}(2013)\citenamefont{Kocsis, Bordacs,
  Varjas, Penc, Abouelsayed, Kuntscher, Ohgushi, Tokura, and
  Kezsmarki}}]{kocsis_2013}
\bibinfo{author}{\bibfnamefont{V.}~\bibnamefont{Kocsis}},
  \bibinfo{author}{\bibfnamefont{S.}~\bibnamefont{Bordacs}},
  \bibinfo{author}{\bibfnamefont{D.}~\bibnamefont{Varjas}},
  \bibinfo{author}{\bibfnamefont{K.}~\bibnamefont{Penc}},
  \bibinfo{author}{\bibfnamefont{A.}~\bibnamefont{Abouelsayed}},
  \bibinfo{author}{\bibfnamefont{C.~A.} \bibnamefont{Kuntscher}},
  \bibinfo{author}{\bibfnamefont{K.}~\bibnamefont{Ohgushi}},
  \bibinfo{author}{\bibfnamefont{Y.}~\bibnamefont{Tokura}}, \bibnamefont{and}
  \bibinfo{author}{\bibfnamefont{I.}~\bibnamefont{Kezsmarki}},
  \bibinfo{journal}{Phys. Rev. B: Condens. Matter}
  \textbf{\bibinfo{volume}{87}}, \bibinfo{pages}{064416}
  (\bibinfo{year}{2013}).

\bibitem[{\citenamefont{Suchomel et~al.}(2012)\citenamefont{Suchomel,
  Shoemaker, Ribaud, Kemei, and Seshadri}}]{Suchomel2012PRB}
\bibinfo{author}{\bibfnamefont{M.~R.} \bibnamefont{Suchomel}},
  \bibinfo{author}{\bibfnamefont{D.~P.} \bibnamefont{Shoemaker}},
  \bibinfo{author}{\bibfnamefont{L.}~\bibnamefont{Ribaud}},
  \bibinfo{author}{\bibfnamefont{M.~C.} \bibnamefont{Kemei}}, \bibnamefont{and}
  \bibinfo{author}{\bibfnamefont{R.}~\bibnamefont{Seshadri}},
  \bibinfo{journal}{Phys. Rev. B: Condens. Matter}
  \textbf{\bibinfo{volume}{86}}, \bibinfo{pages}{054406}
  (\bibinfo{year}{2012}).

\bibitem[{\citenamefont{Ishibashi and Yasumi}(2007)}]{ishibashi_2007}
\bibinfo{author}{\bibfnamefont{H.}~\bibnamefont{Ishibashi}} \bibnamefont{and}
  \bibinfo{author}{\bibfnamefont{T.}~\bibnamefont{Yasumi}},
  \bibinfo{journal}{J. Magn. Magn. Mater.} \textbf{\bibinfo{volume}{310}},
  \bibinfo{pages}{e610} (\bibinfo{year}{2007}).

\bibitem[{\citenamefont{Tomiyasu and Kagomiya}(2004)}]{tomiyasu_2004}
\bibinfo{author}{\bibfnamefont{K.}~\bibnamefont{Tomiyasu}} \bibnamefont{and}
  \bibinfo{author}{\bibfnamefont{I.}~\bibnamefont{Kagomiya}},
  \bibinfo{journal}{J. Phys. Soc. Jpn.} \textbf{\bibinfo{volume}{73}},
  \bibinfo{pages}{2539} (\bibinfo{year}{2004}).

\bibitem[{\citenamefont{Klemme and van Miltenburg}(2002)}]{Klemme_2002}
\bibinfo{author}{\bibfnamefont{S.}~\bibnamefont{Klemme}} \bibnamefont{and}
  \bibinfo{author}{\bibfnamefont{J.~C.} \bibnamefont{van Miltenburg}},
  \bibinfo{journal}{Phys. Chem. Miner.} \textbf{\bibinfo{volume}{29}},
  \bibinfo{pages}{663} (\bibinfo{year}{2002}).

\bibitem[{\citenamefont{Maignan et~al.}(2012)\citenamefont{Maignan, Martin,
  Singh, Simon, Lebedev, and Turner}}]{maignan_2012}
\bibinfo{author}{\bibfnamefont{A.}~\bibnamefont{Maignan}},
  \bibinfo{author}{\bibfnamefont{C.}~\bibnamefont{Martin}},
  \bibinfo{author}{\bibfnamefont{K.}~\bibnamefont{Singh}},
  \bibinfo{author}{\bibfnamefont{C.}~\bibnamefont{Simon}},
  \bibinfo{author}{\bibfnamefont{O.~I.} \bibnamefont{Lebedev}},
  \bibnamefont{and} \bibinfo{author}{\bibfnamefont{S.}~\bibnamefont{Turner}},
  \bibinfo{journal}{J. Solid State Chem.} \textbf{\bibinfo{volume}{195}},
  \bibinfo{pages}{41} (\bibinfo{year}{2012}).

\bibitem[{\citenamefont{Singh et~al.}(2011)\citenamefont{Singh, Maignan, Simon,
  and Martin}}]{singh_2011}
\bibinfo{author}{\bibfnamefont{K.}~\bibnamefont{Singh}},
  \bibinfo{author}{\bibfnamefont{A.}~\bibnamefont{Maignan}},
  \bibinfo{author}{\bibfnamefont{C.}~\bibnamefont{Simon}}, \bibnamefont{and}
  \bibinfo{author}{\bibfnamefont{C.}~\bibnamefont{Martin}},
  \bibinfo{journal}{Appl. Phys. Lett.} \textbf{\bibinfo{volume}{99}},
  \bibinfo{pages}{172903} (\bibinfo{year}{2011}).

\bibitem[{\citenamefont{Toby}(2001)}]{toby_expgui_2001}
\bibinfo{author}{\bibfnamefont{B.}~\bibnamefont{Toby}}, \bibinfo{journal}{J.
  Appl. Crystallogr.} \textbf{\bibinfo{volume}{34}}, \bibinfo{pages}{210}
  (\bibinfo{year}{2001}).

\bibitem[{\citenamefont{Larson and Dreele}(2000)}]{larson_2000}
\bibinfo{author}{\bibfnamefont{A.~C.} \bibnamefont{Larson}} \bibnamefont{and}
  \bibinfo{author}{\bibfnamefont{R.~B.~V.} \bibnamefont{Dreele}},
  \bibinfo{journal}{Los Alamos National Laboratory Report}
  p.~\bibinfo{pages}{86} (\bibinfo{year}{2000}).

\bibitem[{\citenamefont{Detwiler et~al.}(2000)\citenamefont{Detwiler,
  Schmiedeshoff, Harrison, Lacerda, Cooley, and Smith}}]{detwiler_2000}
\bibinfo{author}{\bibfnamefont{J.~A.} \bibnamefont{Detwiler}},
  \bibinfo{author}{\bibfnamefont{G.~M.} \bibnamefont{Schmiedeshoff}},
  \bibinfo{author}{\bibfnamefont{N.}~\bibnamefont{Harrison}},
  \bibinfo{author}{\bibfnamefont{A.~H.} \bibnamefont{Lacerda}},
  \bibinfo{author}{\bibfnamefont{J.~C.} \bibnamefont{Cooley}},
  \bibnamefont{and} \bibinfo{author}{\bibfnamefont{J.~L.} \bibnamefont{Smith}},
  \bibinfo{journal}{Phys. Rev. B: Condens. Matter}
  \textbf{\bibinfo{volume}{61}}, \bibinfo{pages}{402} (\bibinfo{year}{2000}).

\bibitem[{\citenamefont{Catalan}(2006)}]{catalan_2006}
\bibinfo{author}{\bibfnamefont{G.}~\bibnamefont{Catalan}},
  \bibinfo{journal}{Appl. Phys. Lett.} \textbf{\bibinfo{volume}{88}},
  \bibinfo{pages}{102902} (\bibinfo{year}{2006}).

\bibitem[{\citenamefont{Tackett et~al.}(2007)\citenamefont{Tackett, Lawes,
  Melot, Grossman, Toberer, and Seshadri}}]{tackett_2007}
\bibinfo{author}{\bibfnamefont{R.}~\bibnamefont{Tackett}},
  \bibinfo{author}{\bibfnamefont{G.}~\bibnamefont{Lawes}},
  \bibinfo{author}{\bibfnamefont{B.~C.} \bibnamefont{Melot}},
  \bibinfo{author}{\bibfnamefont{M.}~\bibnamefont{Grossman}},
  \bibinfo{author}{\bibfnamefont{E.~S.} \bibnamefont{Toberer}},
  \bibnamefont{and} \bibinfo{author}{\bibfnamefont{R.}~\bibnamefont{Seshadri}},
  \bibinfo{journal}{Phys. Rev. B: Condens. Matter}
  \textbf{\bibinfo{volume}{76}}, \bibinfo{pages}{024409}
  (\bibinfo{year}{2007}).

\bibitem[{\citenamefont{Lawes et~al.}(2003)\citenamefont{Lawes, Ramirez, Varma,
  and Subramanian}}]{lawes_2003}
\bibinfo{author}{\bibfnamefont{G.}~\bibnamefont{Lawes}},
  \bibinfo{author}{\bibfnamefont{A.~P.} \bibnamefont{Ramirez}},
  \bibinfo{author}{\bibfnamefont{C.~M.} \bibnamefont{Varma}}, \bibnamefont{and}
  \bibinfo{author}{\bibfnamefont{M.~A.} \bibnamefont{Subramanian}},
  \bibinfo{journal}{Phys. Rev. Lett.} \textbf{\bibinfo{volume}{91}},
  \bibinfo{pages}{257208} (\bibinfo{year}{2003}).

\bibitem[{\citenamefont{Kimura et~al.}(2003{\natexlab{b}})\citenamefont{Kimura,
  Kawamoto, Yamada, Azuma, Takano, and Tokura}}]{kimura_2003}
\bibinfo{author}{\bibfnamefont{T.}~\bibnamefont{Kimura}},
  \bibinfo{author}{\bibfnamefont{S.}~\bibnamefont{Kawamoto}},
  \bibinfo{author}{\bibfnamefont{I.}~\bibnamefont{Yamada}},
  \bibinfo{author}{\bibfnamefont{M.}~\bibnamefont{Azuma}},
  \bibinfo{author}{\bibfnamefont{M.}~\bibnamefont{Takano}}, \bibnamefont{and}
  \bibinfo{author}{\bibfnamefont{Y.}~\bibnamefont{Tokura}},
  \bibinfo{journal}{Phys. Rev. B: Condens. Matter}
  \textbf{\bibinfo{volume}{67}}, \bibinfo{pages}{180401(R)}
  (\bibinfo{year}{2003}{\natexlab{b}}).

\end{thebibliography}
\end{document}